\documentclass[runningheads]{llncs}

\usepackage{graphicx}
\usepackage{textcomp}
\usepackage{xcolor}
\usepackage{hyperref}
\usepackage{booktabs}
\usepackage{lipsum}
\usepackage{comment}
\usepackage{dependencies}
\usepackage{tabularx}
\usepackage{rotating}
\usepackage{svg}

\usepackage{blindtext}

\usepackage{algorithm,algpseudocode}
\usepackage{amsmath}
\usepackage{amssymb}
\usepackage{cleveref}

\begin{document}

\title{Cross-border Exchange of CBDCs \\
using Layer-2 Blockchain}

\author{Krzysztof Gogol\inst{1}\thanks{Corresponding Author: \texttt{gogol@ifi.uzh.ch}} \and
Johnnatan Messias \inst{2}\and
Malte Schlosser\inst{1}\and
Benjamin Kraner\inst{1}\and
Claudio Tessone\inst{1}
}

\authorrunning{K. Gogol, et al.}

%\author{Anonymous}
%\author{Krzysztof Gogol
%\thanks{gogol@ifi.uzh.ch, Department of Computer Science, University of Zurich}, 
%Johnnatan Messias
%\thanks{Matter Labs}, 
%Malte Schlosser
%\thanks{Department of Banking and Finance, University of Zurich}
%,\\ 
%Benjamin Kraner 
%\thanks{Department of Computer Science, University of Zurich}, 
%and Claudio Tessone
%\thanks{UZH Blockchain Center, University of Zurich}}

\institute{
University of Zurich \and
Matter Labs
}

%\institute{University of Zurich\and
%ETH Zurich
%\email{lncs@springer.com}\\
%\url{http://www.springer.com/gp/computer-science/lncs} \and
%Matter Labs\\
%\email{\{abc,lncs\}@uni-heidelberg.de}}

%
\maketitle              % typeset the header of the contribution
\begin{abstract}
This paper proposes a novel multi-layer blockchain architecture for the cross-border trading of CBDCs. The permissioned layer-2, by relying on the public consensus of the underlying network, ensures the security and integrity of the transactions and interoperability with domestic CBDC implementations. Multiple Layer-3s operate various Automated Market Makers (AMMs) and compete with each other for the lowest costs.
Simulations of trading costs are conducted based on historical FX rates to provide insights into the practical implications of the system, with Project Mariana as a benchmark. The study shows that a multi-layer and multi-AMM setup is more cost-efficient even with liquidity fragmentation than a single AMM.

\end{abstract}

%%%%%%%%%%%%%%%%%%%%%%%%%%%%%%%%%%%%%%%%%%%%%%%%%%%%%%%%%%%%%%%%%%%%%%%%%%%%%%%%%%%%%%%%%%%%%%%%%%%%%%%%%%%%%%%%%%
\section{Introduction}
    \label{sec:intro}

The Bitcoin network, described in a whitepaper by the anonymous individual or group known as Satoshi Nakamoto, launched in 2008 \cite{Nakamoto2008Bitcoin:System}. The fundamental promise of this first cryptocurrency was to establish a peer-to-peer payment system that operates independently of traditional financial intermediaries.
In 2014, Vitalik Buterin introduced Ethereum \cite{Buterin2014Ethereum:Platform.}, the next-generation blockchain with a virtual machine capable of executing computer programs known as smart contracts. With Ethereum and the Ethereum Virtual Machine (EVM), Decentralized Finance (DeFi) was born. By utilizing smart contracts, DeFi offers sophisticated financial services, such as trading, lending, borrowing, derivatives and asset management \cite{werner2022sok}, \cite{schaer2023defimarkets}, \cite{Auer2023technologydefi}, \cite{Gogol2023SoK:Risks}.

It quickly became apparent that this new blockchain-based financial system required non-volatile tokens, giving the advent to stablecoins—tokens with values pegged to fiat currencies. The success of stablecoins prompted central banks to explore similar solutions, leading to the development of Central Bank Digital Currencies (CBDCs)~\cite{ward2019understanding}. While CBDCs do not necessarily have to be implemented on a blockchain, the decentralized approach promises enhanced scalability and security \cite{BIS2023Mariana}, \cite{joitmc7010072}, \cite{app12094488}, \cite{chaum2021issue}. Various central banks across the globe are already engaged in CBDC initiatives\cite{BIS2023Mariana}, \cite{joitmc7010072}, \cite{DREX-Digital-Real-Brazil}. The collective effort explores ways to integrate these CBDCs into a unified marketplace, potentially transcending borders, as seen in  Project Mariana by the Bank for International Settlements (BIS) \cite{BIS2023Mariana}.

This paper introduces a novel architectural framework for cross-border CBDC trading that is based on Layer-2 (L2) blockchain scaling. L2s proved to be an efficient approach for addressing the limitations of Ethereum, especially scalability and privacy \cite{sguanci2021layer}, \cite{yee2022shades}, \cite{gangwal2022survey}. The proposed system is the rollup (non-custodial L2), on the public L1 blockchain that seamlessly integrates additional Layer-3 (L3) blockchains, each operating decentralized exchanges (DEXes), or other DeFi protocols, utilizing CBDCs.  The system automatically selects L3 and DEXes with optimal costs. This system architecture is compared to the approach of Project Mariana in a series of simulations based on the historical Foreign Exchange (FX) rates.

%%%%%%%%%%%%%%%%%%%%%%%%%%%%%%%%%%%%%%%%%%%%%%%%%%%%%%%%%%%%%%%%%%%%%%%%%%%%%%%%%%%%%%%%%%%%%%%%%%%%%%%%%%%%%%%%%%
\section{Background}
    \label{sec:background}
This section introduces the concept of L2 blockchain scaling, presents the functionalities of Project Mariana, and summarizes this work's contributions to the research on cross-border CBDCs exchange.

\subsection{Layer-2 Blockchains}
There are two approaches to tackle the challenges of blockchain scalability and related high gas prices: Layer-1 (L1) and Layer-2 (L2) blockchains. L1 scaling involves creating entirely new blockchains, e.g., with unique consensus mechanisms or block sizes. These chains operate with their own validators and infrastructure, requiring decentralization to win users' trust.
L2 scaling takes a different approach and uses complex computations off-chain (outside of Ethereum), aiming to reduce on-chain data congestion. These off-chain activities include rollups, sidechains, plasma, and state channels. Layer-2 blockchains rely on the security of underlying Layer-1 chains, such as Ethereum, for final settlement.

\begin{figure}[t]
\centering
    \includegraphics[width=0.6\textwidth]{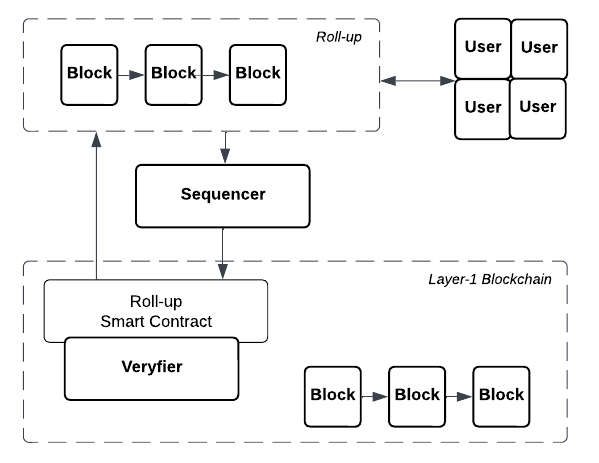} 
\caption{Architecture of Rollup—Layer-2 blockchain that stores state in underlying Layer-1 blockchain}
\label{fig:rollup}
\end{figure}

Rollups represent a form of L2 scaling that does not custody any data by themselves. Rollups offload complex calculations from the Ethereum mainnet and store the results (along with other transactions) in Ethereum after compressing them.
Fig.~\ref{fig:rollup} illustrates rollup architecture with key components—sequencers and verifiers. Sequencers roll up transactions to the Layer-1 chain. By bundling transactions, rollups manage to save on gas fees. Verifiers are smart contracts that operate on Ethereum and verify the transactions stored by the sequencer. They ensure the correctness of the transactions.

%%%%%%%%%%%%%%%%%%%%%%%%%%%%%%%%%%%%%%%%%%%%%%%%%%%%%%%%%%%%%%%%%%%%%%%%%%%%%%%%%%%%%%%%%%%%%%%%%%%%%%%%%%%%%%%%%%
\subsection{Project Mariana}
The specific focus of this work is on facilitating cross-border exchanges of CBDCs. One notable proposal for such a solution is Project Mariana on L1 blockchain (L1-Mariana) \cite{BIS2023Mariana}, initiated by BIS, the Swiss National Bank, the Bank of France, and the Monetary Authority of Singapore.

\begin{figure}[t]
\centering
%\centerline{\includesvg[width=1\linewidth]{diagrams/L2CBDC - Mariana Arch.svg}}
    \includegraphics[width=0.75\textwidth]{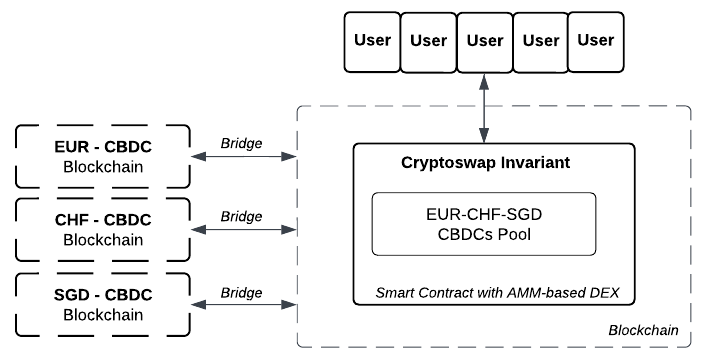} 
\caption{Project Mariana Architecture: Cross-Border CBDC Exchange using Cryptoswap Invariant AMM }
\label{fig:MarianaArchitecture}
\end{figure}
 
Within L1-Mariana, a DEX based on AMM is employed to exchange wholesale CBDCs versions of Swiss Franc (CHF), Euro (EUR), and Singaporean Dollar (SGD). The selected AMM is the Stableswap Invariant, first introduced by Curve v2—the second largest DEX in terms of trading volumes in permissionless DeFi on Ethereum \cite{2022DeFiLlama}. The setup involves bridges that facilitate the transition of wholesale CBDCs from domestic central bank blockchains to the chain where the Cryptoswap Invariant pool operates, as depicted in Fig.~\ref{fig:MarianaArchitecture}. 

Whereas the project proved the feasibility of AMM-DEX for cross-border CBDCs exchange, several questions arise—e.g., the choice of the blockchain (and its scalability and security), the AMM and the pool types (3-token pools vs 2-token pool)—which this research addresses.

%%%%%%%%%%%%%%%%%%%%%%%%%%%%%%%%%%%%%%%%%%%%%%%%%%%%%%%%%%%%%%%%%%%%%%%%%%%%%%%%%%%%%%%%%%%%%%%%%%%%%%%%%%%%%%%%%%
\section{Related Work and Contribution}
%Our study presents a new perspective on the efficiency of multi-AMM setups for CBDC exchanges on Layer 2 blockchains, suggesting they might offer more cost-effective solutions compared to traditional single AMM systems, such as those used in the Mariana project. This is notable even considering the potential challenges of liquidity fragmentation across different AMMs. We introduce a multi-AMM system designed for private Layer 2 blockchains, offering a nuanced approach to scaling blockchain technology. The system's performance was evaluated through simulations based on historical FX rates, with comparisons made to the existing Project Mariana setup. Our contributions include an analysis of the relative cost efficiency of multi-AMM setups in handling diverse transaction volumes and in scenarios of fluctuating gas fees. We provide a detailed breakdown of the total costs involved in CBDC swaps on AMM-DEX, covering aspects like gas fees, swap fees, and slippage costs across various AMMs and blockchain layers. Additionally, we offer quantitative metrics to assess the efficiency of cross-border CBDC transactions on DEXs. Lastly, we have developed a simulation framework that can be applied to other cross-border CBDC projects or within permissioned DeFi contexts in the Ethereum ecosystem, incorporating established mathematical and numerical methods such as the Newton-Raphson method for slippage calculations in the CryptoSwap Invariant AMM context.

There is an abundance of literature discussing the implementation of CBDCs in various countries \cite{joitmc7010072}, \cite{app12094488}, \cite{chaum2021issue}, \cite{MORALESRESENDIZ2021100022}, \cite{DREX-Digital-Real-Brazil}, but little research on the exchange between CBDCs in AMMs, except for Project Mariana \cite{BIS2023Mariana}. Lipton and Sepp\cite{lipton2021automated} study the implicit costs (price impact) for G-10 CBDCs exchange at AMM with Concentrated Liquidity (Uniswap v3). 

This work is the first research that takes the holistic approach to analyze the total costs of CBDCs exchange at various AMMs, including Stableswap Invariant (Curve v2), applied in Project Mariana. This study found that a multi-AMM set-up for CBDCs exchange on L2 is more cost-efficient than a single AMM of Project Mariana, despite the fragmentation of liquidity among various AMMs. Such a multi-AMM system on private L2 is further presented, and its behavior is evaluated via simulation based on the historical FX rates, Mariana Project set up as a benchmark. 

This work makes the following contributions.
\begin{itemize}
\item Demonstrating that multi-AMMs set-up for CBDCs exchange is more cost-efficient than a single AMM.

\item Designing such multi-AMMs set-up for CBDCs exchange using L2 and L3 blockchains and demonstrating its superior performance for small, medium, and largest transaction volumes and any transaction in case of spikes of gas fees.

\item Conducting first detailed analysis of the total costs of CBDCs swaps on AMM-DEX—gas fees, swap fees (explicit fees), and price impact (explicit fees), among various AMMs and L1 and L2-based DEXes.

\item Providing quantitative metrics to evaluate the efficiency of cross-border CBDCs on DEXes for major AMMs. Their implementation is available in the public GitHub repository~\cite{GithubGogol}

\item Building a framework for simulation based on historical FX rate that can be replicated in other cross-border CBDCs projects or in permission DeFi in the Ethereum ecosystem. For all metrics, the mathematical background and numerical methods with the implementation are provided, such as the Newthon-Raphon method for price impact in Cryptoswap Invariant AMM applied in Project Mariana.
\end{itemize}

%%%%%%%%%%%%%%%%%%%%%%%%%%%%%%%%%%%%%%%%%%%%%%%%%%%%%%%%%%%%%%%%%%%%%%%%%%%%%%%%%%%%%%%%%%%%%%%%%%%%%%%%%%%%%%%%%%
\section{System Architecture}
    \label{sec:architecture}

\begin{figure*}[!ht]
  \centering
   % \includesvg[width=0.7\textwidth]{diagrams/L2CBDC - General The System Arch}
        \includegraphics[width=1.1\textwidth]{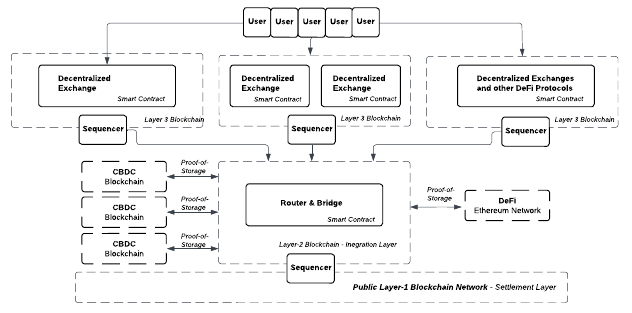} 
    \caption{General Architecture of Cross-Border CBDCs Exchange using L3 and L2 Blockchains on Public L1 Blockchain}
    \label{fig:GeneralTheSystemArch}
\end{figure*}

This paper proposes a novel approach for facilitating cross-border CBDCs exchange through a rollup (a form of L2 blockchain) on the public Layer-1 network, further enhanced with a Layer-3 (L3), as illustrated in Fig.~\ref{fig:GeneralTheSystemArch}. %The Bitcoin network (Layer-1) is used as the settlement and security layer, rollup (Layer-3) is the data integration layer responsible for integration with blockchains hosting CBDCs, and finally, Layer-3s are hosting decentralized exchange and other DeFi protocols that utilize CBDCs. 

\emph{Blockchain and VM Type.}
The system operates as a L2 blockchain on the public L1 network, benefiting from the inherent security and infrastructure of the L1, including consensus mechanisms and miners/validators. Bitcoin, Ethereum or other L1 networks can be the underlying L1. The system also incorporates Virtual Machines (VMs) to execute smart contracts, with the Ethereum Virtual Machine (EVM) being our chosen VM due to its compatibility with most DeFi protocols.

\emph{L2 and L3s Responsibilities.}
L2 is the data integration layer and its operations are managed by central institutions like BIS, focusing on interoperability with other CBDC blockchains and facilitating communication between L2 and L3s.
The system supports multiple independent L3 blockchains, each with the authority to determine the deployment of decentralized exchanges (DEX) and Automated Market Makers (AMM). Furthermore, the deployment of other DeFi protocols is feasible within this framework. AMM and exchange logic of L3s is moved to the external entities that compete with each other.

\emph{Transaction Privacy and Compliance.}
The L2 and L3 blockchains are private (permissioned networks). All participants must undergo KYC~\cite{rajput2013research} procedures before entering the system. This ensures the system can maintain full transaction privacy, e.g. by leveraging zero-knowledge proofs technology~\cite{fiege1987zero}, while adhering to regulatory requirements.

\emph{Proof-of-storage.}
We propose utilizing proof of storage protocols to enhance the security, instead of bridges. Unlike bridges that often involve trust assumptions and centralization, the system employs a proof-of-storage approach. CBDC tokens are directly minted on L2, and proof-of-storage protocols rely on corresponding token reserves within country-specific CBDC blockchains.

\emph{Rule-based router to L3s.}
At the core of the system is a Rule-based router hosted on L2. This router employs rules to select the optimal AMM-DEX and L3 for traders at any given moment. Trading fees are dependent on gas costs and trade expenses, and the router optimizes these parameters for each trade.

\emph{Integration with DeFi on Ethereum.}
To ensure compatibility with regulations, the system integrates with DeFi through the privacy pools on the Ethereum blockchain, proposed by Buterin et al.\cite{buterin2023blockchain}. Only profiles and users validated through these pools are permitted to trade CBDCs, with no additional restrictions or limitations imposed.

\emph{Comparison to Mariana Project.}
A key distinction is its flexibility regarding AMMs, allowing external operators to make decisions about the types of AMMs or central limit order books they wish to operate on L3. Operators of L3s retain control over actions and DeFi protocols within their respective layers. Another advantage of the L2-based system with L3 enhancements is cost optimization and resistance to network congestion. The L2 router automatically directs traders to the L3 AMM-DEX with the lowest trade and gas fees.

%%%%%%%%%%%%%%%%%%%%%%%%%%%%%%%%%%%%%%%%%%%%%%%%%%%%%%%%%%%%%%%%%%%%%%%%%%%%%%%%%%%%%%%%%%%%%%%%%%%%%%%%%%%%%%%%%%
\section{Model}
    \label{sec:model}
The major metric to consider for evaluation of the cross-border CBDCs exchange is the cost of the CBDCs swap.
In general, the total cost of DeFi service consists of three components: \cite{Xu_2023}, \cite{gudgeon2020defi}
\begin{align}
\label{eq:totalcost}
    Total Cost = Gas Fee + DeFi Explicit Fee + DeFi Implicit Fee 
\end{align}

\emph{Gas Fees} are paid to the blockchain operators for executing the transaction. In the case of transactions on roll-ups, the gas fee is comprised of both the L1 and L2 gas fee components.

\emph{DeFi Explicit Fee} is charged by the DeFi protocol for the provided service. In the case of token swap at AMM-DEX, the liquidity pool fee (LP Fee) is the explicit fee charged by the DEX, and part of it (or whole) is distributed to the liquidity providers (LPs) to the pool.

\emph{DeFi Implicit Fee} are indirect fees specific to the design of the DeFi protocol. Price impact and slippage are implicit fees accrued by a trader when swapping tokens at AMM \cite{Xu_2023}, \cite{gudgeon2020defi} and refer to the difference between quoted and executed price by the AMM. Price impact is caused by the transaction volume and slippage (positive or negative) results from other swaps occurring within the same block as the transaction swap.

Further the following model for the total cost of swap is considered:
\begin{align}
\label{eq:swapcost}
Total Swap Cost = Gas Fee + LP Fee + PriceImpact
\end{align}
LP Fee is typically the percentage of the transaction volume, and liquidity providers supply capital to the AMM's liquidity pools in exchange for participation in LP fees. Price impact is the change in exchange price caused by the transaction volume and is specific to the AMM type and the liquidity pool size. The next section presents the major AMM types and corresponding price impact calculation methods.  %In the next section, the total costs of CBDCs exchange are back-tested for the Mariana system and the proposed L2-L3 Exchange.

\subsection{Automated Market Makers}

Constant Function Market Makers are the dominant type of AMMs. They are equipped with a reserve curve function $F: \mathbb{R}^N_+ \rightarrow \mathbb{R}$ that maps $N$ token reserves $x_i$ to a fixed invariant (i.e., constant) $k$. The AMM-DEX in DeFi with the highest trading volumes \cite{2022DeFiLlama} are Curve v2 \cite{Egorov2021CurvePeg} and Uniswap v3 \cite{Adams2021UniswapCore}. The Project Mariana involves Curve v2 AMM and a 3-token pool. The formulation of AMMs follows \cite{Xu2021SoK:Protocols}.

    %\paragraph{Constant Sum} $\sum_{i=1}^N x_i = k$
    %\paragraph{Constant Product} $\prod_{i=0}^N x_i = k$
    %\paragraph{Constant Product with Concentrated Liquidity} $(x_1+\frac{L}{\sqrt{p_b}} )(x_2+L\cdot \sqrt{p_a}) = L^2$
\subsubsection*{Constant Product with Concentrated Liquidity Market Maker (CLMM)  - Uniswap v3}
    \begin{equation}\label{eq:uniswapv3}(x_1+\frac{L}{\sqrt{S_l}} )(x_2+L\cdot \sqrt{S_u}) = L^2\end{equation}

\subsubsection*{Stableswap Invariant  - Curve v1} \begin{equation}\label{eq:curvev1}
        K\cdot D^{N-1}\cdot \sum\limits_{i=1}^N x_i +\prod\limits_{i=1}^N x_i = K\cdot D^N + (\frac{D}{N})^N
    \end{equation} where $K= A\cdot\prod_{i=1}x_i \cdot D^{-N} \cdot N^N$.

\subsubsection*{Cryptoswap Invariant - Curve v2} As in equation (\ref{eq:curvev1}), but with K defined as \begin{equation}\label{eq:curvev2}K = A\cdot \underset{\text{ =: }K_0}{\underbrace{\frac{A\cdot\prod_{i=1}x_i}{D^N}\cdot N^N}} \cdot \frac{\gamma^2 }{(\gamma+1- K_0)^2}\end{equation}
\subsection{Price Impact}

Further, we consider 2-token, or 3-token pools and denote the reserves of CBDCs in the pools by $x_t$ for CHF, $y_t$ for EUR, and $z_t$ for SGD at time t. The FX spot rate CHF-EUR is $S_t$ and CHF-SGD is $P_t$. The token reserves $x_0$, $y_0$, $z_0$ at the moment of the pool creation are set by:
\begin{align}
\label{eq:startPoint}
x_0 = N_0, y_0 =S_0 N_0, z_0= P_0 N_0
\end{align}
For a buy transaction of $\Delta y$ EUR in exchange for $\Delta x$ CHF at time t and the slippage is:
\begin{align}
PriceImpact(\Delta x, \Delta y, S_t) = \frac{\Delta y / \Delta x}{S_t} - 1 
\end{align}
%\paragraph{Slippage costs at 2-token pool at CPMM}
%\begin{align}
%Slippage(\Delta y, S_t, S_0, N_0) = \frac{\Delta y}{y_t} = \sqrt{ \frac{S_t}{S_0}} \frac{\Delta x}{N_0}
%\end{align}
%\begin{align}
%S_t = \frac{y_t}{x_t} \\
%\end{align}
%\emph{Slippage Costs at Concentrated Liquidity}
In CLMM, LPs can define the price range $[S_l, S_u]$, for which liquidity is provided. When the exchange price $S_t$ moves outside the range $[S_l, S_u]$, LP does not earn any fees. For simplicity, we consider the range $[\frac{S_t}{\alpha},  S_t \alpha ]$, where $\alpha > 1$. Slippage at CLMM can be expressed as \cite{Xu2021SoK:Protocols}:
\begin{align}
\label{eq:slippageCLMM}
PriceImpact(t) =\frac{\Delta y}{N_0*\sqrt{ S_0 * S_t}} (1-\frac{1}{\sqrt{\alpha}})
\end{align}
%\emph{Slippage Costs at Cryptoswap Invariant} 
For $n>3$, the closed formula for slippage at Cryptoswap Invariant does not exist \cite{tiruviluamala2022general}. The numerical approximation of the Newton-Raphson method is used in two steps. First, given Eq.~\ref{eq:startPoint}, the parameter D is calculated, and, subsequently, $\Delta x$ is found.

%%%%%%%%%%%%%%%%%%%%%%%%%%%%%%%%%%%%%%%%%%%%%%%%%%%%%%%%%%%%%%%%%%%%%%%%%%%%%%%%%%%%%%%%%%%%%%%%%%%%%%%%%%%%%%%%%%
%%%%%%%%%%%%%%%%%%%%%%%%%%%%%%%%%%%%%%%%%%%%%%%%%%%%%%%%%%%%%%%%%%%%%%%%%%%%%%%%%%%%%%%%%%%%%%%%%%%%%%%%%%%%%%%%%%
\section{Simulation}
    \label{sec:simulation}

To ensure comparability with L1-Mariana , we consider three CBDCs: Swiss Franc (CHF), Euro (EUR), and Singaporean Dollar (SGD) and back-test the total costs of CBDCs swaps in L1-Mariana and L2-L3-Exchange systems. We consider the buy transaction of EUR for CHF for volumes ranging from 1 EUR to 1mn EUR. The price impact for CLMMs are calculated based on Eq.~\ref{eq:slippageCLMM}, and the Newton-Raphson method is applied for Cryptoswap Invariants, both for $n=2$ and $n=3$. The implementation of all methods is available in a public GitHub repository~\cite{GithubGogol} and was based on the smart contract code of Curve v2. 

\subsection{Data and Parameters}
 The analysis uses the historical exchange rates of CHF-EUR and CHF-SGD for the last three years (daily closing values from Yahoo Finance). We assume that arbitrageurs maintain prices aligned between CBDCs in AMMs and the FX market, which allows us to calculate the historical AMM pools' compositions based on these FX rates.

\emph{Liquidity pools.}
We assume 100mn CHF is provided to each system for liquidity provisions. As L1-Mariana operates Cryptoswap Invariant with one 3-token pool CHF-EUR-SGD, the pool's value is 100mn CHF. 
In L2-L3-Exchange, we consider three L3 operators, each running AMM-DEX setup. The 100mn CHF liquidity is equally divided among the L3 operators. 
The first L3 operator uses a Cryptoswap Invariant with CHF-EUR-SGD pool (like in L1-Mariana), and the pool is valued 100/3 mn CHF. The second L3 also employs Cryptoswap Invariant, but with two 2-token pools: CHF-EUR and CHF-SGD, each valued 100/6 mn CHF. The third L3 operator utilizes CLMM with CHF-EUR and CHF-SGD pools, each valued 100/6 mn CHF. The L2-L3-Exchange pool architecture is illustrated in Fig.~\ref{fig:TheSystemArch} in the appendix.

\emph{AMMs.} To be compatible with L1-Mariana, for all Cryptoswap Invariant we set A to 50 and gamma to $10^{-8}$. For CLMM, we consider the liquidity range around the spot price with the alpha of 1.2, rebalanced daily.

\emph{Swap fees.} Each liquidity pool charges swap fee of 1bp (i.e., 0.0001) of transaction volume.

\emph{Gas costs.} Gas fees for swaps on Ethereum, referred to in the Project Mariana, oscillated between 5 EUR and 50 EUR over the last three years \cite{2023Ethererscan}. The gas costs on L2s are, on average, 50 times lower compared to the underlying L1 \cite{gangwal2022survey}, \cite{yee2022shades}. Based on these facts, we consider 15 EUR gas fees for a swap on L1 and 15EUR/50 on the L2-L3 network. 

%%%%%%%%%%%%%%%%%%%%%%%%%%%%%%%%%%%%%%%%%%%%%%%%%%%%%%%%%%%%%%%%%%%%%%%%%%%%%%%%%%%%%%%%%%%%%%%%%%%%%%%%%%%%%%%%%%
\subsection{Results}
%    \label{sec:analysis}

\begin{figure}[!thb]
    \centering
    \includegraphics[width=1\textwidth]{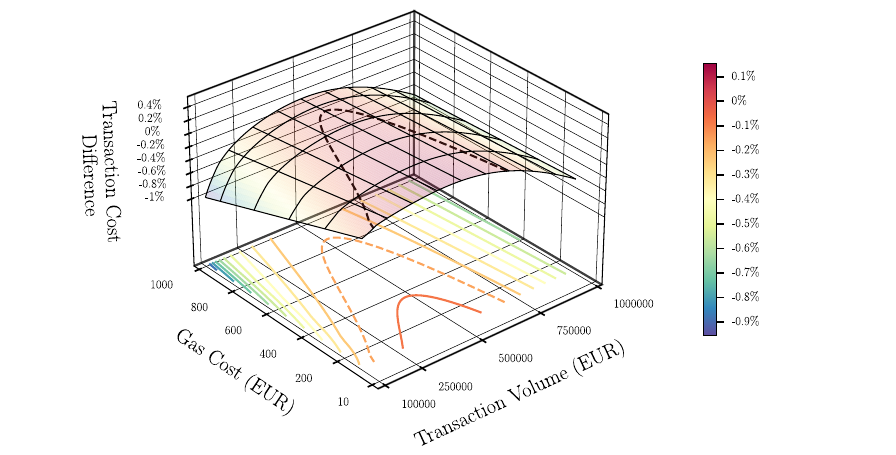} 
    \caption{Difference between the total swap costs at L1-Mariana and L2-L3-Exchange. The negative difference indicates L2-L3-Exchange out-performance for small and large transaction volumes.} 
    \label{fig:3D}
\end{figure}

\begin{figure}[!thb]
    \centering
    \includegraphics[width=1.1\linewidth]{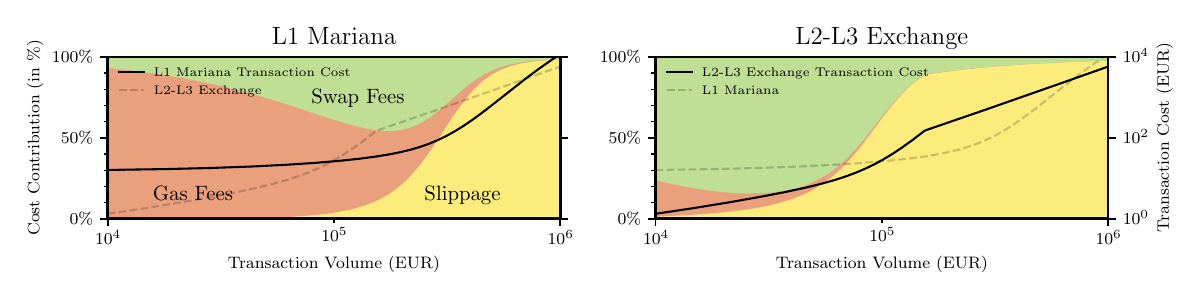} 
    \caption{Breakdown of total swap costs into gas fees, LP fees (swap fee) and price impact (slippage) as a function of the swap volume} 
    \label{fig:costBreakdown}
\end{figure}

\begin{figure}[!thb]
  \begin{minipage}[c]{0.49\linewidth}
    \includegraphics[width=1\linewidth]{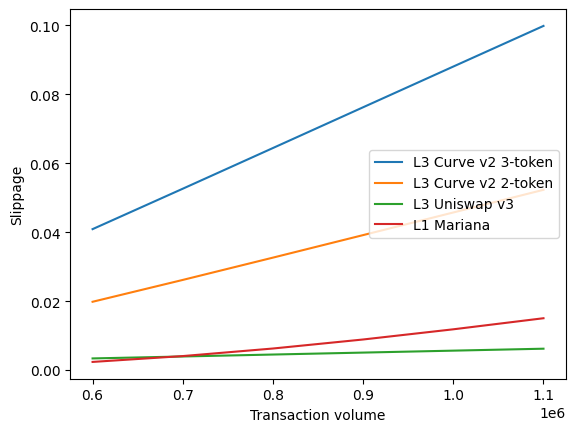} 
  \end{minipage} 
    \begin{minipage}[c]{0.49\linewidth}
    \includegraphics[width=1\linewidth]{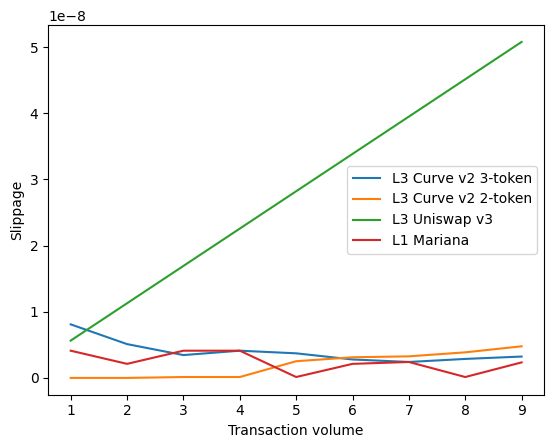} 
  \end{minipage} 
  \caption{Price impact at L1 Mariana and L3 AMM operators. The liquidity pool at L1 Mariana AMM is worth 100mn EUR and each pool of L3 AMM is worth 33mn EUR} 
  \label{fig:slippage}
\end{figure}

The simulation outcomes are depicted in Figs.~\ref{fig:3D}, \ref{fig:costBreakdown} and \ref{fig:slippage}. Fig. \ref{fig:TotalCostCharts} in the appendix presents the historical total swap costs for transaction volumes of 10k, 100k, and 1mn EUR for CHF for each day over the last 3 years with fixed gas fees. Fig \ref{fig:3D} extends this analysis for various gas fees and presents the difference between the costs of L2L3 Exchange and L1 Mariana. Notably, L2-L3-Exchange outperforms for volumes of 10k and 1mn. This outperformance varies, influenced not only by transaction volume but also by current gas fees, as shown in \cref{fig:3D}. For gas fees surpassing 800 EUR, L2-L3-Exchange outperforms for any transaction volume. For lower gas fees, such as 15 EUR, the out performance range is observed for volumes representing less than 0.01\% or more than 0.7\% of the provided liquidity.
The reason for L2-L3-Exchange outperforming for smaller transactions lies in the minimization of gas fees through the application of L2 and L2s. Fig. \ref{fig:costBreakdown} breaks down total swap costs, indicating that gas fees dominate the costs for L1 Mariana in smaller transactions, compared to the swap fee for L2-L3-Exchange. For larger transactions, price impact becomes the predominant component of total swap costs.

The outperformance of the L2-L3-Exchange over L1-Mariana and its ability to handle price impact effectively stem from the use of L3 with CLMM, as illustrated in Figs. \ref{fig:slippage}. The analysis of L3 selection indicates the use of L3 with Cryptoswap Invariant 2-token pool for smaller transactions, while the 3-token pool of Cryptoswap Invariant is unused for any transaction volumes. L3 with CLMM and 2-token pool Cryptoswap Invariant outperforms, in certain situations, the 3-token pool of CryptoSwap Mariana despite liquidity fragmentation, with each L3 having three times less liquidity than L1 Mariana.
Fig. \cref{fig:costBreakdown} explains the role of gas, swap, and price impact components in overall transaction costs. %Based on these findings, the appropriate AMM can be selected to achieve optimal results. 
The simulation showed that applying more AMMs, despite the liquidity fragmentation, outperforms the application of just one AMM: 100mn to L1-Mariana and 33.33mn to each L3. Fig. \ref{fig:tvl_contour} simulates the swap costs when various total liquidity is provided. 

%a) In the case of large transactions, L2-L3 Exchange outperforms Mariana. This superiority arises from the involvement of Uniswap v3 in L3.
%b) For small transactions (below xx), both systems exhibit zero slippage costs. However, due to lower gas fees in L2-L3, this system emerges as the preferable choice.

%c) When gas fees on L1 surpass 800 EUR, L2-L3 Exchange consistently outperforms Mariana across all transaction volumes.

%d) L3 comparisons:

%The 3-token pool is not utilized.
%The 2-token pool performs well up to xxx volume, beyond which Uniswap v3 becomes more effective.
%e) Gas fees dominate in the domain of small volumes with swap fees, whereas slippage takes precedence in large volumes.

%f) The overall proportion of gas fees in L2-L3 Exchange is smaller than in L1.

\begin{table*}[ht]
\centering
\caption{Breakdown of the average fees for the small (10k EUR), medium (100k EUR) and large (1mn EUR) transactions. The buy EUR-CHF transactions are considered at L1 Mariana and L2-L3 Exchange. All fees are denominated in EUR. The simulation is based on the daily historical FX rate for the last three years.}
\begin{tabular}{l llll}
\hline
                   & Total Fee & Gas Fee & Swap Fee & Price Impact \\
         Size   &           &         &          &          \\
\hline
  \textit{L1 Mariana:}             &           &         &          &          \\
                Small  & 16.0 (16bps) & 15.0 (15bps) & 1.0 (1bps) & 0.01 (0bps) \\
                Medium & 26.1 (3bps) & 15.0 (2bps) & 10.0 (1bps) & 1.08 (0bps)         \\
                Large  & 10,386.4 (104bps) & 15.0 (0bps) & 100.0 (1bps) & 10,271.42 (103bps)       \\
    \textit{L2 Exchange:}       &           &         &          &          \\
                Small  & 1.3 (1bps) & 0.3 (0bps) & 1.0 (1bps) & 0.02 (0bps)         \\
                Medium & 25.6 (3bps) & 0.3 (0bps) & 10.0 (1bps) & 15.33 (2bps)        \\
                Large  & 5,445.3 (54bps)& 0.3 (0bps) & 100.0 (1bps) &  5,345.04 (52bps)       \\ \hline
\end{tabular}
\end{table*}
%%%%%%%%%%%%%%%%%%%%%%%%%%%%%%%%%%%%%%%%%%%%%%%%%%%%%%%%%%%%%%%%%%%%%%%%%%%%%%%%%%%%%%%%%%%%%%%%%%%%%%%%%%%%%%%%%%
\section{Discussion}
    \label{sec:discussion}
    
%\emph{Private vs Public Blockchain} 
The primary consideration when establishing cross-border CBDCs exchange is the choice of blockchain architecture. While private L1s may initially appear to provide greater operational control, they require significant investments in physical infrastructure. Public blockchains operate in a decentralized manner, supported by well-established consensus mechanisms. A novel approach involves \emph{permissioned L2s} networks on public blockchains, which combine the advantages of both private and public networks.

Another system decision is about allowance arbitrage, where the revenue from arbitrage could be allocated to AMM-DEX operators, L2-L3 operators or central banks. There is ongoing research about sequencer decentralization at public L2 blockchains and the enablement of MEV and MEV arbitrage. Currently, Ethereum roll-ups operations are supported only by one sequencer which does not allow MEV.

To assess the system, additional metrics such as latency and finality can be considered in conjunction with the domestic CBDC blockchains and the underlying L1 network. In this context, \emph{latency} refers to the duration between submitting a transaction to a network and receiving the initial confirmation of its commitment. On the other hand, \emph{finality} refers to the time required for confirming that a transaction is irrevocable.
%Other metrics to evaluate the cross-border CBDCs exchange system might include latency and finality with the domestic CBDC blockchains and with the underlying Layer-1 network. \emph{Latency} is the time between submitting a transaction to a network and the first confirmation of acceptance by the network, and \emph{finality} is the time until the confirmation that a transaction is unchangeable. 
%\emph{Environmental Aspects} Bitcoin network relies on Proof of Work (PoW), a mechanism that consumes electricity to ensure network security. There are currently around 1mn miners worldwide consuming 127 (TWh) a year, which is more than many countries. The Ethereum blockchain, on the other hand, employs Proof of Stake (PoS) consensus, which is environmentally friendly. The network's security is upheld by 27.3M staked Ether (worth ca 50bn\$) and maintained by over 0.8mn validator. Given the emphasis on sustainability and decentralization, it is prudent to consider building the Layer-2 blockchain for CBDC trading on the Ethereum platform. 
%\emph{Censorship Resistance} Other public blokchains did not achieve the decentralization and security levels of Bitcoin and Ethereum. The alternative approach to developing rollups on Bitcoin and Ethereum is the plasma chain. In such solutions, the cross-border CBDCs exchange takes place on (public or private) layer-1 blockchain, but the state of all transactions is saved in the Bitcoin and Ethereum networks, allowing any participants to publicly validate the correctness of transactions.
%\emph{Future Research}

Furthermore, the profitability of liquidity provisions and corresponding swap fee, should be analyzed. The prototype implementation of such systems could be based on the ready rollup frameworks \cite{2022ZkSync}, \cite{2022Starknet}, \cite{2023Polygon}.

%%%%%%%%%%%%%%%%%%%%%%%%%%%%%%%%%%%%%%%%%%%%%%%%%%%%%%%%%%%%%%%%%%%%%%%%%%%%%%%%%%%%%%%%%%%%%%%%%%%%%%%%%%%%%%%%%%
\section{Conclusions}

The major finding of this research is that a multi-AMM setup on L2 for a CBDCs exchange is more cost-efficient than a single L1 AMM, despite the liquidity fragmentation. This work presented the design of such a multi-AMM system on private L2.
The proposed system offers the advantages of private blockchain while minimizing infrastructure costs and enhancing security, as it leverages the consensus mechanism of the underlying L1. 

Through a historical simulation based on the FX rates among CHF, EUR, and SGD, we compared the exchange costs between L2-L3 Exchange and Project Mariana. We found that the L2-L3 Exchange outperforms across small, low-medium, and large transactions, even in the presence of liquidity fragmentation between L3s. During gas fee spikes, the performance gap widens, with lower transaction costs for all transactions at L2-L3 Exchange. Our analysis of CBDC swap costs revealed that the gas fees component is significantly lower in the L2-L3 Exchange for medium transactions, enabling their efficient swap.

%\newpage

\section*{Acknowledgements} 
The first author is a research fellow at Matter Labs. The second author is a researcher at Matter Labs. This research article is a work of scholarship and reflects the authors' own views and opinions. It does not necessarily reflect the views or opinions of any other person or organization, including the authors' employer. Readers should not rely on this article for making strategic or commercial decisions, and the authors are not responsible for any losses that may result from such use.

\bibliographystyle{splncs04}
\bibliography{main}

%%
%% If your work has an appendix, this is the place to put it.
\newpage
\appendix

%%%%%%%%%%%%%%%%%%%%%%%%%%%%%%%%%%%%%%%%%%%%%%%%%%%%%%%%%%%%%%%%%%%%%%%%%%%%%%%%%%%%%%%%%%%%%%%%%%%%%%%%%%%%%%%%%%
\section*{Graphs and Tables}

\begin{comment}
\begin{figure*}[]
  \centering
    \includesvg[width=\textwidth]{diagrams/L2CBDC - Mariana Arch.svg}
    \caption{Architecture of Cross-Border CBDC Exchange in Project Mariana}
    \label{fig:MarianaArchitecture}
\end{figure*}

\begin{figure*}[]
  \centering
    \includesvg[width=\textwidth]{diagrams/L2CBDC - Rollup}
    \caption{Architecture of Rollup}
    \label{fig:rollup}
\end{figure*}
\end{comment}

\begin{figure*}[!h]
  \centering
    %\includesvg[width=\textwidth]{diagrams/L2CBDC_TheSystemArch}
    \includegraphics[width=1\linewidth]{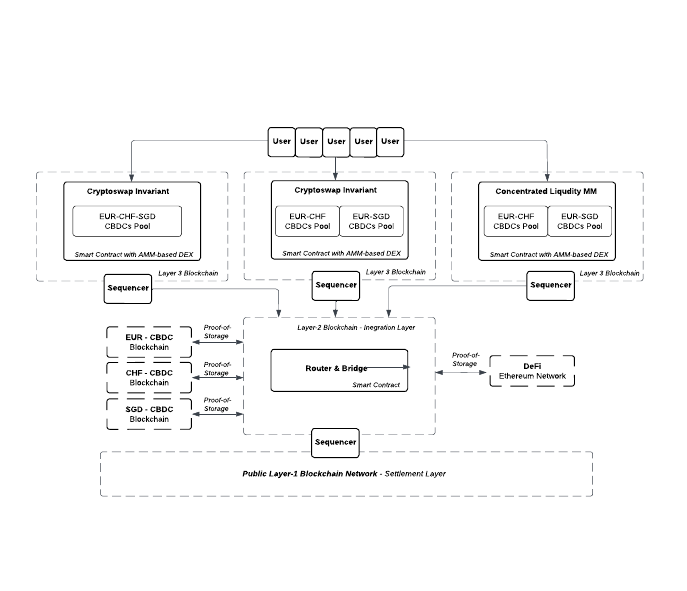} 
    \caption{Architecture of Cross-Border CBDC Exchange using L3 and L2s on the Public Layer-1 Blockchain Network for EUR, CHF and SGD CBDCs. The selected exchanges are Cryptoswap Invariant with one pool of three CBDCs, Cryptoswap Invariant with two pools of two CBDCs and Concentrated Liquidity Market Maker with two pools of two CBDCs}
    \label{fig:TheSystemArch}
\end{figure*}

\newpage

%%%%%%%%%%%%%%%%%%%%%%%%%%%%%%%%%%%%%%%%%%%%%%%%%%%%%%%%%%%%%%%%%%%%%%%%%%%%%%%%%%%%%%%%%%%%%%%%%%%%%%%%%%%%%%%%%%
%\section*{Slippage Costs for Major AMMs}
%    \label{sec:AMMs}

\begin{comment}
We denote the EURCHF FX spot rate by $p_t$ and assume that $p_0$ is the rate at the pool creation with notional $p_0 N_0$ EUR and $N_0$ CHF. The initial pool balances in EUR, $x_0$,
and CHF $y_0$ are set respectively by:
\begin{align}
x_0 = p_0 N_0, y_0 = N_0.
\end{align}

The buy and sell prices of tokens from the AMM pool are determined by the CFMM function:
\begin{align}
F(x_1; y_1; x_0; y_0, k) = 0;
\end{align}
where $x_1$ and $y_1$ are the amounts of EUR and CHF tokens after a transaction, respectively; $x_0$ and $y_0$ are the amounts right before the transaction, $k$ is constant. 

Buying $\Delta x$ EUR tokens from the pool involves deposing $\Delta y$ tokens to the CHF pool. The AMM applies the CFMM to determine the amount of $\Delta x$ in exchange for $\Delta y$ as follows:
\begin{align}
F(x_0 - \Delta x; y_0 + \Delta y) = 0
\end{align}
\end{comment}

\begin{figure}[ht]
    \begin{minipage}[c]{0.33\linewidth}
        \includegraphics[width=0.99\linewidth]{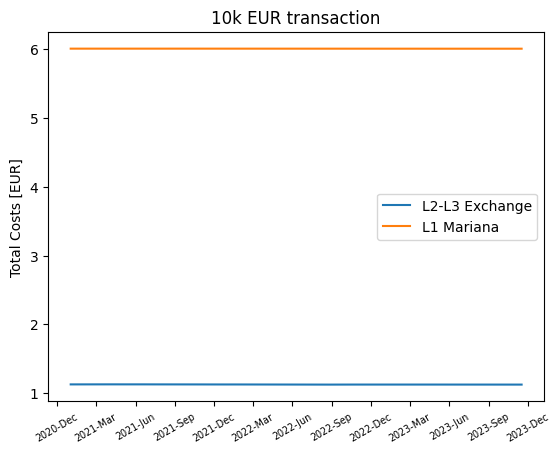} 
    \end{minipage}%
    \begin{minipage}[c]{0.33\linewidth}
        \includegraphics[width=0.99\linewidth]{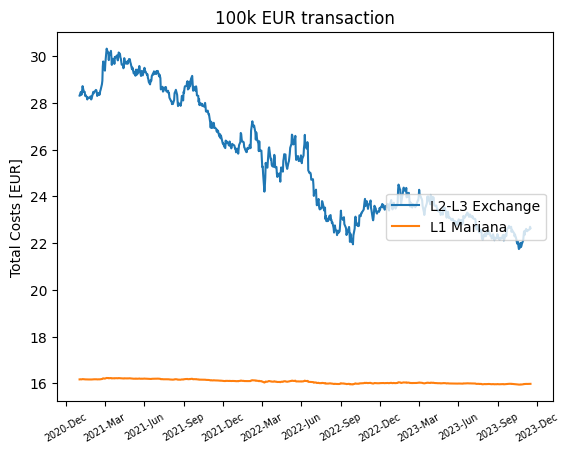} 
    \end{minipage}%
    \begin{minipage}[c]{0.33\linewidth}
        \includegraphics[width=0.99\linewidth]{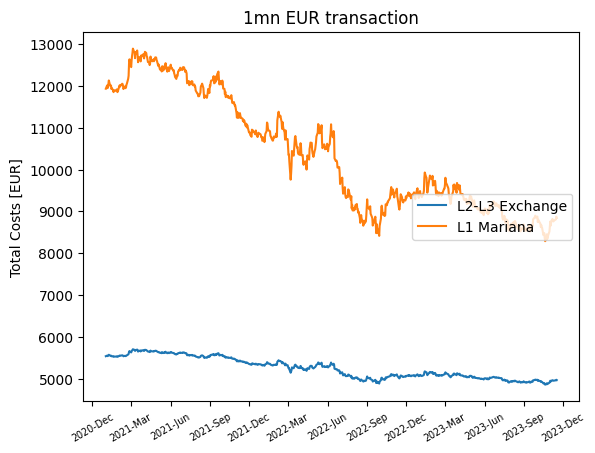} 
    \end{minipage}%
    %\caption{Total costs of exchange EUR CBDC to CHF CBDCs based on historical exchange rates} 
    %\label{fig:DVTCharts} 
    %\label{fig:dailyCosts}
  \hfill
  \begin{minipage}[c]{0.33\linewidth}
    \includegraphics[width=0.99\linewidth]{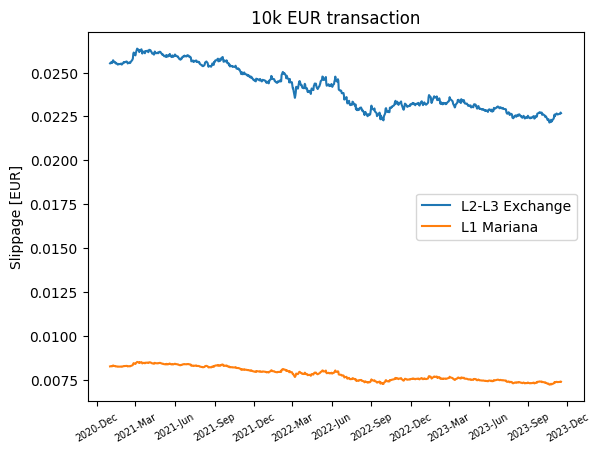} 
  \end{minipage}%
  \begin{minipage}[c]{0.33\linewidth}
    \includegraphics[width=0.99\linewidth]{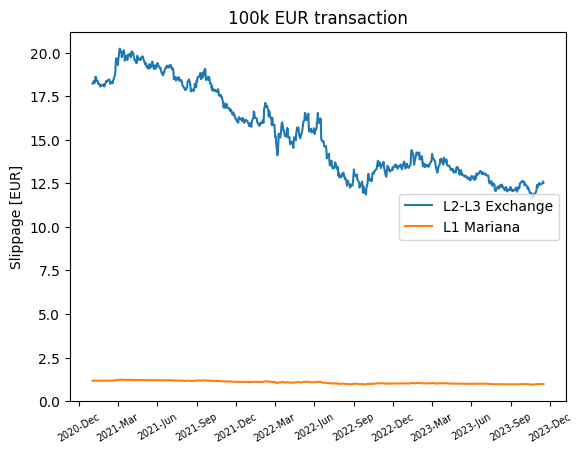} 
  \end{minipage}%
  \begin{minipage}[c]{0.33\linewidth}
    \includegraphics[width=0.99\linewidth]{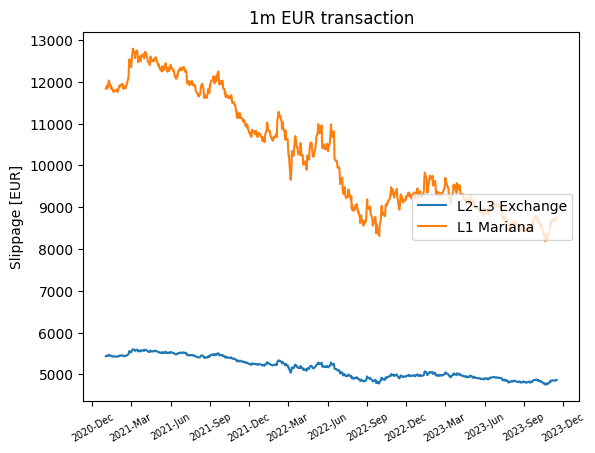} 
  \end{minipage}%
  \hfill
  \begin{minipage}[c]{0.33\linewidth}
    \includegraphics[width=0.99\linewidth]{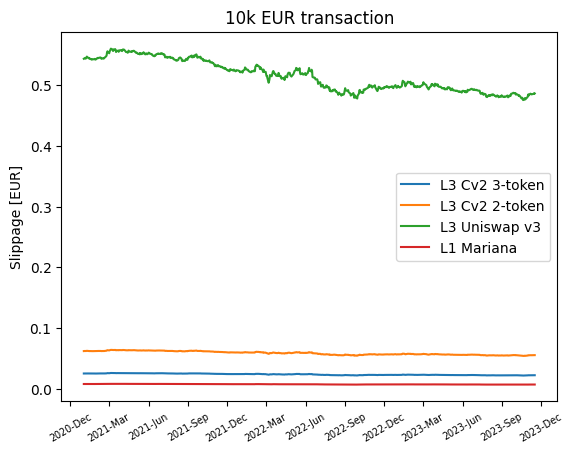} 
  \end{minipage}%
  \begin{minipage}[c]{0.33\linewidth}
    \includegraphics[width=0.99\linewidth]{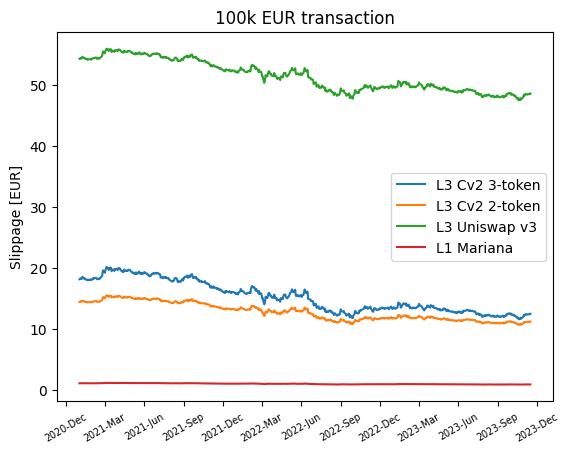} 
  \end{minipage}%
  \begin{minipage}[c]{0.33\linewidth}
    \includegraphics[width=0.99\linewidth]{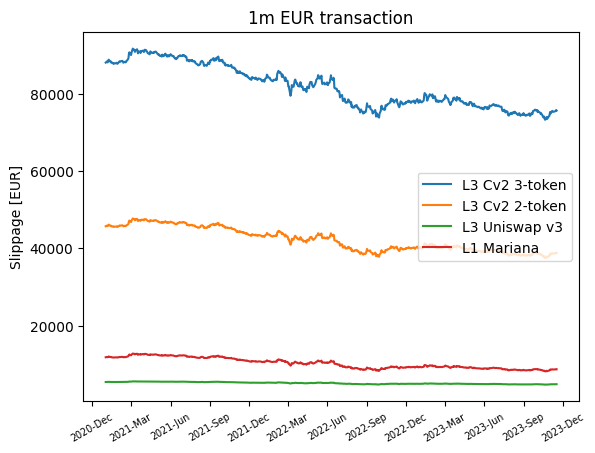} 
  \end{minipage}%
  \caption{Back-testing results for transaction volumes of 10k, 100k and 1mn, assuming 15 EUR gas fees and 100mn TVL provided to each system (based on the historical FX rates)} 
  \label{fig:TotalCostCharts} 
\end{figure}

%\begin{figure}[!t]
%    \centering
%    \includegraphics[width=\textwidth]{diagrams/cost_breakdown.pdf} 
%    \caption{Evolution of total costs breakdown for CBDCs exchange} 
%    \label{fig:cost_breakdown}
%\end{figure}

\begin{figure}[!t]
    \centering
    \includegraphics[width=\textwidth]{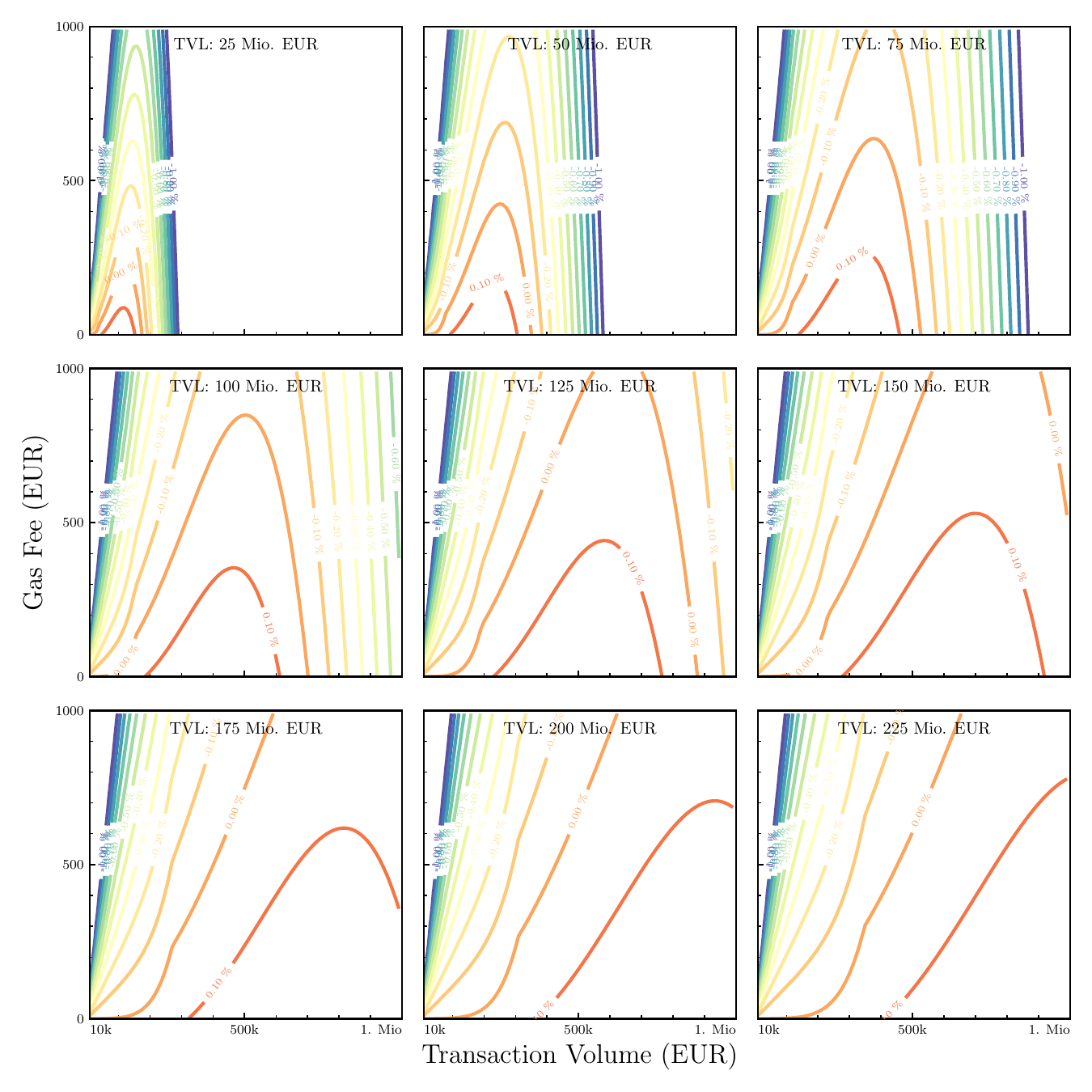} 
    \caption{The contour lines of the percentage difference (of total transaction cost) between L1-Mariana and an L2-L3 exchange for different size of market. The Total Value Locked is indicated in each panel. \cref{fig:3D}, refers to the left most panel, in the middle row.} 
    \label{fig:tvl_contour}
\end{figure}

\end{document}